\documentclass[twocolumn,showpacs,preprintnumbers,amsmath,amssymb]{revtex4}

\usepackage{graphicx}
\usepackage{dcolumn}
\usepackage{bm}

\begin{document}


\title{A note on the equivalence of Lagrangian and Hamiltonian formulations at post-Newtonian approximations} 
\author{Rongchao Chen}
\author{Xin Wu}
\email{xwu@ncu.edu.cn}
\affiliation{Department of
Physics and Institute of Astronomy, Nanchang University, Nanchang 330031, China}


\begin{abstract}

It was claimed recently that a low order post-Newtonian (PN) Lagrangian formulation,
which corresponds to the Euler-Lagrange equations up to an infinite PN order, can be identical to
a PN Hamiltonian formulation at the infinite order from a theoretical point of view.
This result is difficult to check because in most cases one does not know what
both the Euler-Lagrange equations and the equivalent Hamiltonian
are at the infinite order. However, no difficulty exists for a special 1PN
Lagrangian formulation of relativistic circular restricted three-body problem,
where both the Euler-Lagrange equations and the equivalent Hamiltonian not only
are expanded to all PN orders but also have converged functions. Consequently,
the analytical evidence supports this claim. As far as numerical evidences are concerned,
the Hamiltonian equivalent to the Euler-Lagrange equations for the lower order Lagrangian
requires that they both be only at higher enough finite orders.

\end{abstract}

\pacs{04.25.Nx, 05.45.-a, 45.20.Jj, 95.10.Fh}

\maketitle

\section{Introduction}

In classical mechanics, Lagrangian and Hamiltonian formulations are completely the same
description of a dynamical system.
Usually more attention to the Hamiltonian formulation is paid
because it has properties of a canonical system.

In post-Newtonian (PN) mechanics of general relativity, the two formulations are still adopted. Are they completely equivalent?
Ten years ago two independent groups [1,2] answered this question. They proved the complete physical equivalence of
the third-order post-Newtonian (3PN) Arnowitt-Deser-Misner (ADM) coordinate
Hamiltonian approach to and the 3PN harmonic coordinate Lagrangian approach to
the dynamics of spinless compact binaries. This result was recently extended to the inclusion of the next-to-next-to-leading order (4PN)
spin-spin coupling [3].

However, there are two different claims on the chaotic behavior of compact binaries
with one body spinning and spin effects restricted to spin-orbit (1.5PN) coupling. That is,
the 2PN harmonic coordinate Lagrangian dynamics allow the onset of chaos [4],
but the 2PN ADM Hamiltonian dynamics
are integrable, regular and non-chaotic [5,6].

An explanation to the opposite results was given in [7].
In fact, the 2PN Hamiltonian and  Lagrangian formulations are not exactly equal but are only approximately
related. As its detailed account, the equations of motion for the Lagrangian formulation use lower-order terms as approximations to
higher-order acceleration terms
in the Euler-Lagrange equations, while these approximations do not occur in the equations of motion for the Hamiltonian formulation.
It is natural that the Lagrangian has approximate constants of motion but the Hamiltonian contains exact ones. These facts were regarded as
the essential point for the two formulations having different dynamics. In this sense, the two claims that seem to
be explicitly conflicting were thought to be correct.

Recently, the authors of [8] revisited the equivalence between the Hamiltonian and Lagrangian formulations at PN approximations.
They found that the two formulations at the same PN order are nonequivalent in general and have differences.
Three simple examples of PN Lagrangian formulations, including a relativistic restricted three-body problem with the
1PN contribution from the circular motion of two primary objects,
a spinning compact
binary system with the Newtonian term and the leading-order
spin-orbit coupling [8] and a binary system of the Newtonian term and the leading-order spin-orbit and spin-spin
couplings [9], were used to show that the differences are not mainly due to the Lagrangian
having the approximate Euler-Lagrange equations and the approximate constants of motion but come from
truncation of higher-order PN terms between the two formulations transformed. An important result from the logic is that
an equivalent Hamiltonian of a lower-order Lagrangian is usually at an infinite order from a theoretical point of view or
at a higher enough order from numerical computations. Based on this, the integrability or non-integrability
of the Lagrangian can be known by that of the Hamiltonian. More recently, chaos in
comparable mass compact binary systems with one body spinning was completely ruled out [10]. The reason is that a completely canonical higher-order
Hamiltonian, which is equivalent to a lower-order conservative Lagrangian and holds
four integrals of the total energy and
the total angular momentum in an eight-dimensional phase space,
is typically integrable [11]. This result is useful to clarify the doubt on the absence of chaos in the 2PN ADM Hamiltonian approach [5,6]
and the presence of chaos in the 2PN harmonic coordinate Lagrangian formulation [4]. As a point to illustrate, two other doubts about
different chaotic indicators resulting
in different dynamical behaviours of spinning compact binaries among references [12-15] and different descriptions of chaotic parameter
spaces and chaotic regions between two articles [4,16] have been clarified in [17-19].

It is worth noting that the logic result on the equivalence of the PN Hamiltonian and Lagrangian approaches at different orders
is not easy to check because the exactly equivalent Hamiltonian of the Lagrangian is generally expressed as an infinite series
whose convergence is unknown clearly in most cases. To provide enough evidence for supporting this result,
we select a part of the 1PN Lagrangian formulation
of relativistic circular restricted three-body problem [20], where the Euler-Lagrange equations can be described by a converged Taylor series
and the equivalent Hamiltonian can also be written as another converged Taylor series.
For our purpose, the Hamiltonian is derived from the Lagrangian in Sect. 2. Then in Sect. 3 numerical methods are used to evaluate
whether various PN order Hamiltonians and the 1PN Lagrangian with various PN order Euler-Lagrange equations are equivalent.
Finally, the main results are concluded in Sect. 4.

\section{Post-Newtonian approximations}

As in classical mechanics, a Lagrangian formulation $L(\mathbf{\dot{r}},\mathbf{r})$ and
its Hamiltonian formulation $H(\mathbf{p},\mathbf{r})$ satisfy the Legendre transformation
in PN mechanics. This transformation is written as
\begin{equation}
H(\mathbf{p},\mathbf{r}) = \mathbf{p}\cdot\mathbf{\dot{r}}-L(\mathbf{\dot{r}},\mathbf{r}).
\end{equation}
Here $\mathbf{r}$ and $\mathbf{\dot{r}}$ are coordinate and velocity, respectively.
Canonical momentum is
\begin{equation}
\mathbf{p} = \frac{\partial L(\mathbf{\dot{r}},\mathbf{r})}{\partial \mathbf{\dot{r}}}.
\end{equation}
Taking a special PN circular restricted three-body problem as an example, now we
derive the Hamiltonian from the Lagrangian in detail.

\subsection{Lagrangian formulation}

The circular restricted three-body problem means
the motion of a third body (i.e. a small particle of negligible mass) moving around two masses $m_1$ and $m_2$ ($m_1\geq m_2$).
The two masses move in circular, coplanar orbits about their common center of mass,
and have a constant separation $a$ and the same angular velocity. They
exert a gravitational force on the particle but the third body does not affect
the motion of the two massive bodies. Taking the unit of mass $G(m_1+m_2)=1$, we have the two masses
$\mu_1=m_1/(m_1+m_2)$ and $\mu_2=m_2/(m_1+m_2)$. The unit of length requires that the constant separation
of the two bodies should be unity.
The common mean motion, the Newtonian angular velocity $\omega_0$, of the two primaries is also unity.
In these unit systems, the two bodies are stationary at points $O_1(x_1,0)$ and $O_2(x_2,0)$ with $x_1=-\mu_2$ and $x_2=\mu_1$
in the rotating reference frame. State variables $(\mathbf{\dot{r}},\mathbf{r})$ of
the third body satisfy the following Lagrangian formulation
\begin{eqnarray}
L &=& L_0+\frac{1}{c^{2}}L_1+\frac{1}{c^{2}}L_2, \\
L_0 &=&
\frac{1}{2}(\dot{x}^2+\dot{y}^2+x^2+y^2)+x\dot{y}-\dot{x}y+U,
\\
L_1 &=& \omega_1(x\dot{y}-\dot{x}y +x^2+y^2),
\end{eqnarray}
\begin{eqnarray}
L_2 = \frac{3}{2a}U
[\dot{x}^2+\dot{y}^2+x^2+y^2+2(x\dot{y}-\dot{x}y)].
\end{eqnarray}
In the above equations, the related notations are specified as follows.
$U$ is of the form
\begin{equation}
U=\frac{\mu_1}{r_1}+\frac{\mu_2}{r_2},
\end{equation}
where the distances from body 3 to bodies 1 and 2 are
\begin{eqnarray}
r_{1} &=& \sqrt{(x-x_1)^{2}+y^2}, \nonumber \\
r_{2} &=& \sqrt{(x-x_2)^{2}+y^2}. \nonumber
\end{eqnarray}
$L_0$ stands for the Newtonian circular restricted three-body problem.
$L_1$ is a 1PN contribution due to the relativistic
effect to the circular motions of the two primaries. $L_2$ is also a 1PN contribution
from the relativistic
effect to the third body, and is only a part of that in [20] for our purpose.
$\omega_1$ is the 1PN effect with respect to the angular velocity $\omega_0$ of the primaries
and is given by
\begin{eqnarray}
\omega_1 = (\mu_1\mu_2-3)/(2a).
\end{eqnarray}
In fact, the separation $a$ is a mark of $L_1$ and $L_2$ as the 1PN effects when
the velocity of light, $c$, is taken as one geometric unit in later numerical computations.

The Lagrangian (3) is a function of velocities and coordinates, therefore, its equations of motion are
the ordinary Euler-Lagrange equations:
\begin{eqnarray}
\frac{d}{dt}(\frac{\partial L}{\partial\dot{x}}) =
\frac{\partial L}{\partial x}, ~~~~~~~~
\frac{d}{dt}(\frac{\partial L}{\partial\dot{y}}) =
\frac{\partial L}{\partial y}.
\end{eqnarray}
Since the momenta $p_x=\partial L/\partial\dot{x}$ and $p_y=\partial L/\partial\dot{y}$
of the forms
\begin{eqnarray}
p_{x} &=&
\dot{x}-(1+\frac{\omega_1}{c^{2}})y+\frac{3U}{ac^{2}}(\dot{x}-y),\\
p_{y} &=&
\dot{y}+(1+\frac{\omega_1}{c^{2}})x+\frac{3U}{ac^{2}}(\dot{y}+x)
\end{eqnarray}
are linear functions of velocities $\dot{x}$ and $\dot{y}$,
accelerations can be solved exactly from Eq. (9).
They have
detailed expressions:
\begin{eqnarray}
\ddot{x} &=& \frac{X_0+X_1/c^2}{1+3U/(ac^2)},\\
\ddot{y} &=&\frac{Y_0+Y_1/c^2}{1+3U/(ac^2)}.
\end{eqnarray}
The Newtonian terms $X_0$ and $Y_0$ and the 1PN terms $X_1$ and $Y_1$ are
\begin{eqnarray}
X_0 &=&
x+2\dot{y}+U_x, \\
Y_0 &=&
y-2\dot{x}+U_y,
\end{eqnarray}
\begin{eqnarray}
X_1 &=& 2\omega_1(x+\dot{y})+\frac{U_x}{U}L_2+\frac{3}{a}[U(x+2\dot{y}) \nonumber \\
& & -(\dot{x}U_x+\dot{y}U_y)(\dot{x}-y)], \\
Y_1 &=& 2\omega_1(y-\dot{x})+\frac{U_y}{U}L_2+\frac{3}{a}[U(y-2\dot{x}) \nonumber \\
& & -(\dot{x}U_x+\dot{y}U_y)(x+\dot{y})],
\end{eqnarray}
where $U_x=\partial U/\partial x$ and $U_y=\partial U/\partial y$.
Considering that $\delta=3U/a$ is at the 1PN level, Eqs. (12) and (13)
have the Taylor expansions
\begin{eqnarray}
\ddot{x} &\approx& X_0[\sum\limits_{i=0}^{k}(-1)^{i}(\frac{\delta}{c^2})^{i}]
+\frac{X_1}{c^2}[\sum\limits_{j=0}^{k-1}(-1)^{j}(\frac{\delta}{c^2})^{j}],\\
\ddot{y} &\approx& Y_0[\sum\limits_{i=0}^{k}(-1)^{i}(\frac{\delta}{c^2})^{i}]
+\frac{Y_1}{c^2}[\sum\limits_{j=0}^{k-1}(-1)^{j}(\frac{\delta}{c^2})^{j}].
\end{eqnarray}
They are the Euler-Lagrange equations with PN approximations to an order $k\geq 1$, labeled as
$EL_k$. As a point to illustrate, the case of $k=0$ with $X_1=Y_1=0$ corresponds to the Newtonian Euler-Lagrange equations, marked as
$EL_0$.  From a theoretical viewpoint, as $k\rightarrow\infty$, $EL_k$ is strictly equivalent to $EL$
given by Eqs. (12) and (13), namely, $EL_{\infty}\equiv EL$. Note that for the generic case in [8],
the momenta are highly nonlinear functions of velocities, so no exact equations of motion similar to Eqs. (12) and (13)
but approximate equations of motion can be obtained from the Euler-Lagrange equations (9). This means that we do not know
what the PN approximations like Eqs. (18) and (19) are converged as $k\rightarrow\infty$.

\subsection{Hamiltonian formulations}

The velocities $\dot{x}$ and $\dot{y}$ obtained from Eqs. (10) and (11) are expressed as
\begin{eqnarray}
\dot{x} &=& \frac{p_{x}}{1+\delta/c^2}+(1+\frac{\omega_1}{c^{2}}){y},\\
\dot{y} &=& \frac{p_{y}}{1+\delta/c^2}-(1+\frac{\omega_1}{c^{2}}){x}.
\end{eqnarray}
Of course, they can be expanded to the $k$th order
\begin{eqnarray}
\dot{x} &\approx& p_{x}[\sum\limits_{i=0}^{k}(-1)^{i}(\frac{\delta}{c^2})^{i}]+(1+\frac{\omega_1}{c^2}){y},\\
\dot{y} &\approx& p_{y}[\sum\limits_{i=0}^{k}(-1)^{i}(\frac{\delta}{c^2})^{i}]-(1+\frac{\omega_1}{c^2}){x}.
\end{eqnarray}
As mentioned above, Eqs. (22) and (23) are exactly identical to Eqs. (20) and (21) when $k\rightarrow\infty$.

In light of Eqs. (1), (20) and (21), we have the following Hamiltonian
\begin{eqnarray}
H &=& \frac{1}{2(1+\delta/c^{2})}(p^{2}_{x}+p^{2}_{y})
+(1+\frac{\omega_1}{c^2})(yp_{x}-xp_{y})\nonumber\\
&&-U.
\end{eqnarray}
Its Taylor series at the $k$th order is of the form
\begin{eqnarray}
H_k &=& \frac{1}{2}(p^{2}_{x}+p^{2}_{y})\sum\limits_{i=0}^{k}(-1)^{i}(\frac{\delta}{c^{2}})^{i}
+(1+\frac{\omega_1}{c^2})(yp_{x}-xp_{y})\nonumber\\
&& -U.
\end{eqnarray}
It is clear that $H_0$ with $\omega_1=0$ is the Newtonian Hamiltonian formulation, and can be
expressed in terms of the Jacobian constant $C_J$ as $H_0\equiv-C_J/2$. Additionally,
$H_k$ is closer and closer to $H$ as $k$ gets larger. Without doubt, the exact equivalence between $H$ and $H_k$
should be $H_{\infty}\equiv H$. Of course, what $H_k$ is converged as $k\rightarrow\infty$ is still unknown for the general case in [8].

It should be emphasized that $EL_{k}$ is the $k$th order PN
approximation to the Euler-Lagrange equations $EL$ that is exactly derived from the
1PN Lagrangian $L$, and $H_{k}$ is the $k$th order PN
approximation to the Hamiltonian $H$. Because of the exact equivalence between $EL$ and $H$,
$EL_{k}$ is the $k$th order PN
approximation to the Hamiltonian $H$, and $H_{k}$ is the $k$th order PN
approximation to the Euler-Lagrange equations $EL$.
Additionally, $EL_{\infty}$ and $H_{\infty}$ are exactly equivalent,
i.e., $EL_{\infty}\equiv EL \Leftrightarrow H\equiv H_{\infty}$.
However, it would be up to a certain higher enough finite order $k$ rather than up to the infinite order $k$
that the equivalence $EL_{k}\Leftrightarrow H_{k}$ can be checked by numerical methods. See the following
numerical investigations for more details.

\section{Numerical investigations}

Besides the above analytical method, a numerical method is used to estimate whether these PN approaches
have constants of motion and what the accuracy of the constants is. Above all,
we are interested in knowing whether these PN approaches are equivalent.

\subsection{Energy errors}

An eighth- and ninth-order
Runge--Kutta--Fehlberg algorithm of variable time-steps is used to solve each of
the above Euler-Lagrange equations $EL_k$ and Hamiltonians $H_k$. Parameters and initial conditions
are $C_J=3.12$, $\mu_2=0.001$, $x=y=0.55$ and $\dot{x}=0$. Note that the initial positive value of $\dot{y}$
is given by the Jacobian constant. This orbit in the Newtonian problem $L_0$
is a Kolmogorov-Arnold-Moser (KAM) torus on the Poincar\'{e}
section $y=0$ with $\dot{y}>0$ in Fig. 1(a), therefore, it is regular and non-chaotic. The
integrator can give errors of the energy $H_0$ for the Lagrangian $L_0$ in the magnitude of
order $10^{-13}$ or so. The long-term accumulation of energy errors is explicitly present in Fig. 1(b) because
the integration scheme itself yields an
artificial excitation or damping. If this accumulation is neglected, the energy should be constant.
This shows that the energy $H_0$ is actually an integral of the Lagrangian $L_0$.
However, the existence of this excitation or damping does not make the numerical
results unreliable during the integration time of $10^{5}$ due to such a high numerical accuracy.
In this sense, not only the integrator does not necessarily use manifold correction methods [21-23],
but also it gives true qualitative
results as a symplectic integration algorithm [24-27] does.

When the PN terms $L_1$ and $L_2$ are included, what about the accuracy of energy integrals
given by the related PN approximations? Let us answer this question. Taking the separation
between the primaries, $a=31$, we plot
Fig. 2(a) in which the errors of energies of the 1PN Euler-Lagrange equations $EL_1$ and Hamiltonian
$H_1$ are shown. It is worth noting that the error of energy
is estimated by means of $\Delta=H_1-\tilde{H}_1$, where $H_1$ is regarded as the energy of $EL_1$ at time $t$
and $\tilde{H}_1$ is the initial energy. Obviously, the error for $EL_1$ is larger in about 10 orders of magnitude
than that for $H_1$. This result should be very reasonable because differences between $EL_1$ and $H_1$
exist explicitly but the canonical equations are exactly given by the 1PN Hamiltonian $H_1$, as
shown in the above analytical discussions. In other words, the difference between $EL_1$ and $H_1$
is at 1PN level. Of course, the higher the order $k$ gets, the smaller
the difference between $EL_k$ and $H_k$ becomes. This is why we can see from Figs. 2(a) and 2(b) that
the error of the 8PN Euler-Lagrange equations $EL_8$ and Hamiltonian
$H_8$ is typically smaller than that of the 1PN Euler-Lagrange equations $EL_1$ and Hamiltonian
$H_1$. Without doubt, $EL$ and $H$ should be the same in the energy accuracy if
no roundoff errors exist in Fig. 2(c).

In addition to evaluating the accuracy of energy integrals of these PN approaches, evaluating
the quality of these PN approaches to the Euler-Lagrange equations $EL$ or the Hamiltonian $H$
is also necessary from qualitative and quantitative numerical comparisons.
See the following demonstrations for more information.

\subsection{Qualitative comparisons}

Besides the method of Poincar\'{e}
sections, the method of Lyapunov exponents is often used to
detect chaos from order. It relates to the description of average
exponential deviation of two nearby orbits. Based on the two-particle method [28], the largest
Lyapunov exponent is calculated by
\begin{eqnarray}
\lambda = \lim_{t\rightarrow\infty}\frac{1}{t}\ln\frac{d(t)}{d(0)},
\end{eqnarray}
where $d(0)$ and $d(t)$ are distances between the two nearby trajectories at times 0 and $t$, respectively.
A globally stable orbit is said to be regular if $\lambda=0$ but chaotic if $\lambda>0$.
Generally speaking, it costs a long enough time to obtain a stabilizing value of $\lambda$ from the limit.
Instead, a quicker method to find chaos is a fast Lyapunov indicator [29,30], defined as
\begin{eqnarray}
FLI = \log_{10} \frac{d(t)}{d(0)}.
\end{eqnarray}
The globally stable orbit is chaotic if this indicator increases exponentially with time $\log_{10} t$
but ordered if this indicator grows polynomially.

It can be seen clearly from the Poincar\'{e} section of Fig. 3(a) that the
dynamics of $EL$ or $H$ in Fig. 2(c) is chaotic. This result is supported by the
Lyapunov exponents in Figs. 3(b) and 3(c) and the FLIs in Fig. 3(d) and 3(e). What about the dynamics of
these various PN approximations? The key to this question can be found in Figs. 3(b)-3(e). Here are the related details.
As shown in Fig. 3(b), lower order PN approximations to the Euler-Lagrange equations $EL$, such as
the 1PN Euler-Lagrange equations $EL_1$ and the 4PN Euler-Lagrange equations $EL_4$, are so poorer that
their dynamics are regular, and are completely unlike the chaotic dynamics of $EL$. With increase of the PN order $k$,
higher order PN approximations to the Euler-Lagrange equations $EL$ become better and better.
For example, the 8PN Euler-Lagrange equations $EL_8$ allows the onset of chaos, as $EL$ does.
Seen particularly from
the evolution curve on the Lyapunov exponent and time,
the 12PN Euler-Lagrange equations $EL_{12}$ seems to be very closer to $EL$. These results
are also suitable for the PN Hamiltonian approximations to the Hamiltonian $H$ in Fig. 3(c).
When the Lyapunov exponents in Figs. 3(b) and 3(c)
are replaced with the FLIs in Figs. 3(d) and 3(e), similar results can be given.

When the separation $a=138$ is instead of $a=31$ in Fig. 3(a), an ordered KAM torus occurs.
That means that the $EL$ dynamics is regular and non-chaotic. In Figs 3(f)-3(i), lower order PN approximations such as
$EL_8$ (or $H_8$) have chaotic behaviors, but higher order PN approximations such as
$EL_{12}$ (or $H_{12}$) have regular behaviors.

In short, the above numerical simulations seem to tell us that
the Euler-Lagrange equations (or the Hamiltonian approaches) at higher enough PN orders have the same dynamics as
the Euler-Lagrange equations $EL$ (or the Hamiltonian $H$). There is a question of whether these results depend on the separation $a$.
To answer it, we fix the above-mentioned orbit but let $a$ begin at 10 and end at 250 in increments of 1. For each given value of $a$,
the FLI is obtained after integration time $t=3500$. In this way, we have dependence of FLIs on the separations $a$
in several PN Lagrangian and Hamiltonian approaches, plotted in Fig. 4. Here 5.5 is referred as a threshold value of FLI for
distinguishing between the regular and chaotic cases at this time. That is to say, an orbit is chaotic when its FLI is larger than
threshold but ordered when its FLI is smaller than
threshold. In light of this, we do not find that there are dramatic dynamical differences between the Euler-Lagrange equations $EL$ (or the Hamiltonian $H$)
and the various PN approximations such as the 1PN Hamiltonian
$H_1$ and the 1PN Euler-Lagrange equations $EL_1$. However, it is clearly shown in Table 1 that regular and chaotic domains of
smaller separations $a$ in the lowest PN approaches $EL_1$ and $H_1$ are explicitly different from those in $EL$ or $H$.
As claimed above, this result is of course expected. When the order $k$ gets higher and higher,
$EL_k$ and $H_k$ have smaller and smaller dynamical differences compared with $EL$ or $H$. Two points are worth noting. First,
the same order PN approaches like $EL_{12}$ and $H_{12}$ (but unlike $EL$ and $H$) are incompletely equivalent in the dynamical
behaviors for smaller values of $a$. Second, all the PN approaches $EL_1$,  $H_1$, $EL_{12}$, $H_{12}$, $\cdots$, $EL$ and $H$ can still have the same
dynamics when $a$ is larger enough. The two points are due to
the differences among these approaches from the relativistic effects depending on $a$; smaller values of $a$
result in larger relativistic effects but larger values of $a$
lead to smaller relativistic effects.

\subsection{Quantitative comparisons}

Now we are interested in quantitative studies on the various PN
approximations $EL_{k}$ to the Hamiltonian $H$ and the various PN
approximations $H_{k}$ to the Euler-Lagrange equations $EL$. In other words, we want to know how the
deviation $|\Delta \mathbf{r}|=|\mathbf{r}_k-\mathbf{r}_H|$ between the position coordinate $\mathbf{r}_k$ for $EL_{k}$ (or $H_{k}$)
and the position coordinate $\mathbf{r}_H$ for $H$ (or $EL$) varies with time. To provide
some insight into the rule on the deviation with time, we should consider the regular dynamics in various PN
approximations because the chaotic case gives rise to exponentially sensitive dependence on initial conditions.
For the sake of this purpose, the parameters and initial conditions unlike the aforementioned ones are
$C_J=2.07$, $x=0.68$ and $y=0$. When $a=140$ is given in Fig. 5(a), the curve $EL$ is used to estimate
the accuracy of numerical solutions between $H$ and $EL$, which begins in about the magnitude of $10^{-14}$
and is in about the magnitude of $10^{-7}$ at time $t=10000$.
The difference numerical solutions between $H$ and $EL_1$ is rather large.
With increase of $k$, $EL_{k}$ is soon closer to $H$. For instance, $EL_{8}$ is basically consistent with $H$
after time $t=3000$, and $EL_{12}$ is almost the same as $H$. Similarly, this rule is suitable for
the approximations $H_{k}$ to the Euler-Lagrange equations $EL$ in Fig. 5(b). After the integration time
reaches 10000 for each $a\in[10,10000]$ in Figs. 5(c) and 5(d), appropriately larger separation $a$ and higher enough order $k$
are present such that $EL_{k}$ and $H_{k}$ are identical to $H$ or $EL$.
In a word, it can be seen clearly from Fig. 5 that $EL_{k}$ and $H_{k}$ are equivalent as $k$ is sufficiently large.

\section{Summary}

In general, PN Lagrangian and Hamiltonian formulations at the same order are nonequivalent due to
higher order terms truncated. A lower order Lagrangian is possibly identical to a higher enough order Hamiltonian.
It is difficult to check this equivalence because the Euler-Lagrange equations are not exactly but approximately
derived from the Lagrangian. To cope with this difficulty, we
take a simple relativistic circular restricted three-body problem as an example
and investigate the equivalence of PN Lagrangian and Hamiltonian formulations.
This dynamical problem is described by a 1PN Lagrangian formulation, in which the Euler-Lagrange equations not only
are exactly given but also can be expressed as a converged infinite PN order Taylor series.
The Lagrangian has an exactly equivalent Hamiltonian, expanded to another converged infinite PN order Taylor series.
Numerical results support the equivalence of the 1PN Lagrangian with the Euler-Lagrange equations at a certain specific higher order
and the PN Hamiltonian approach to a higher enough order. In this way, we support indirectly
the general result of [8,10] that a lower order Lagrangian approach
with the Euler-Lagrange equations at some sufficiently higher order
can be equivalent to a higher enough order Hamiltonian approach.

\begin{acknowledgements}
This research is supported by the Natural Science Foundation of Jiangxi Province and the National Natural
Science Foundation of China under Grant Nos. 11173012,
11178002 and 11533004.
\end{acknowledgements}

\begin{table*}
\center{ \caption{Ordered and chaotic domains of the separation $a\in[10,250]$ in Fig.4}
\begin{tabular}{ccccc}
\hline
 System &  Order & Chaos \\
 \hline
$EL_{1}$ &17, [23,25], [27,30], [34,48], 50, 52&[10,16],[18,22],26,[31,33],49,51,[53,57],
\\
&[58,60],[63,84],89,91,[93,99],[129,135]&61,62,[85,88],90,92,[100,128],[136,140]\\
&141,156,[159,173],195,196,[219,250]&[142,155],157,158,[174,194],[197,218]\\
\hline
$H_{1}$ &16, [18,23], [26,43], [49,53], 83, 84, 194 &[10,15],17,24,25,[44,48],54,55,80,81,165
\\
&[88,94], [125,132], 138, 139, 152, 178 & 82,[85,87],[95,124],[133,137],[140,151] \\
&[157,164], [166,170], 195, [219,250] & [153,156],[171,177],[179,193],[196,218]\\
 \hline
$EL_{12}$ & 21, 22, [24,28], [32,46], 48, 137, 138 & [10,20], 23, 29, 30, 31, 47, [194,213]
\\
 & [54,57],[60,80],85,87,[89,95],[125,131]&[49,53],58,59,[81,84],86,88,[96,124] \\
 & [155,169], 191, 192, 193, 214, [216,250]&[132,136], [139,154], [170,190], 215\\
\hline
$H_{12}$ & 17, 21, 22, [24,28], [32,46], 48, 137, 138 &[10,16],18,19,20,23,29,30, [194,213]
\\
 & [54,57], [60,80], 85, 87, [89,95], [125,131] &[49,53],58,59,[81,84],86,88,[96,124],31 \\
 & [155,169], 191, 192, 193, 214, [216,250] & [132,136],[139,154],[170,190],215,47 \\
\hline
$EL$($H$) & [10,12], [14,16], 21, 22, [24,28], [32,46], 48 & 13, [17,20], 23, 29, 30, 31, 47, 215
\\
&[54,57],[60,80],85,87,[89,95],[125,131],137&[49,53],58,59,[81,84],86,88,[96,124] \\
&138,[155,169],191,192,193,214,[216,250]&[132,136],[139,154],[170,190],[194,213] \\
\hline
\end{tabular}}
 \label{tab1}
\end{table*}

\begin{figure*}
\center{
\includegraphics[scale=0.68]{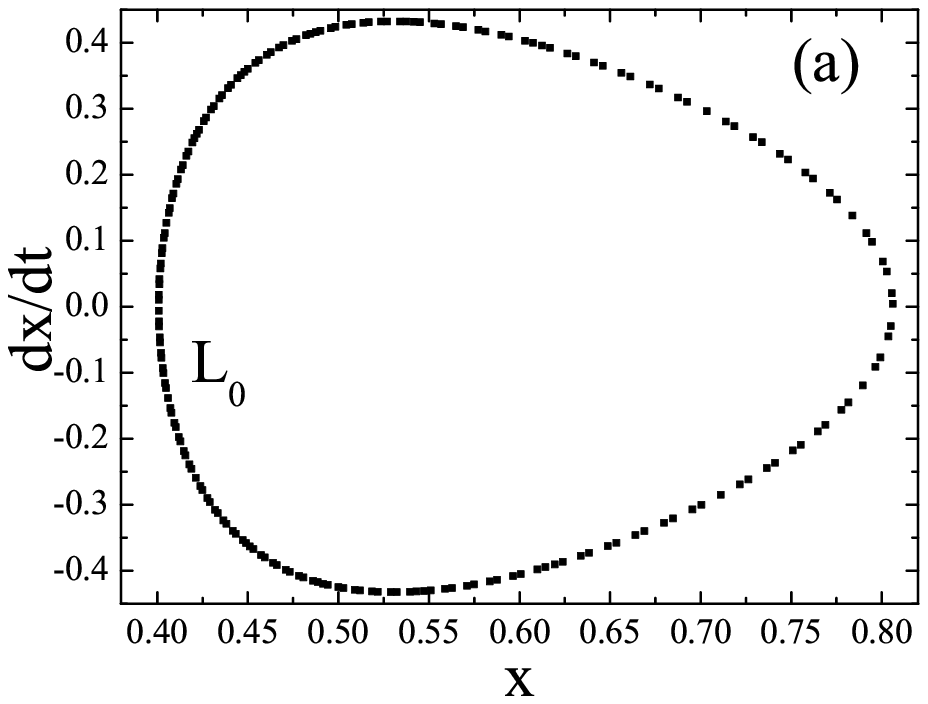}
\includegraphics[scale=0.68]{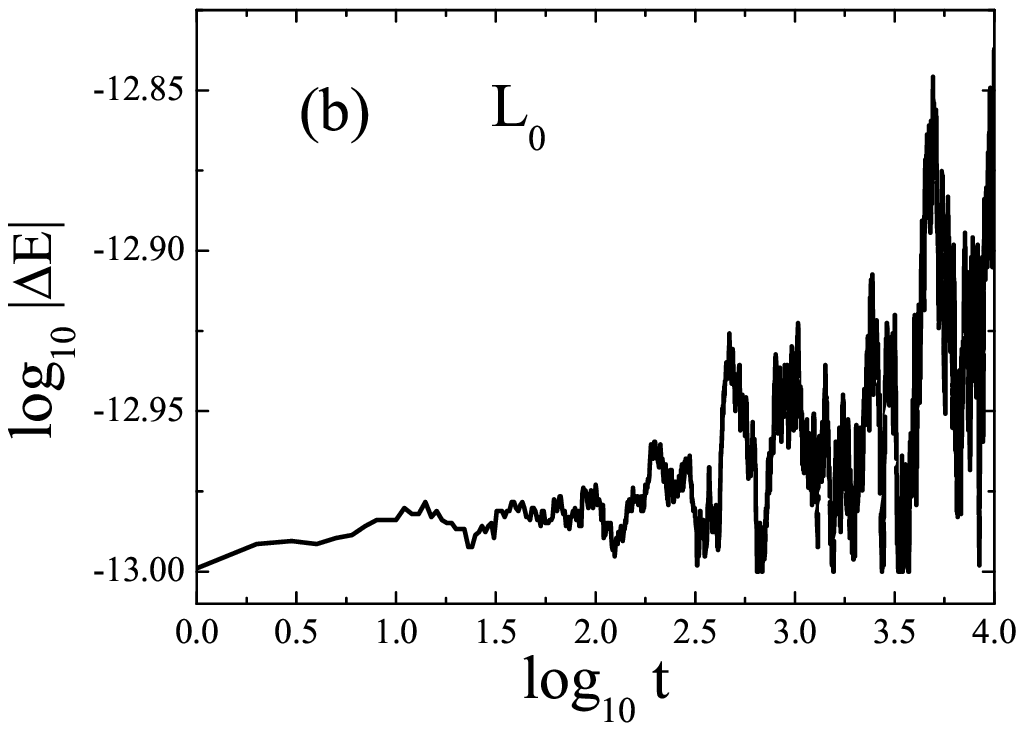}
\caption{Panel (a) Poincar\'{e} section $y=0$ ($\dot{y}>0$) of an orbit with parameters $C_J=3.12$ and $\mu_2=0.001$
and initial conditions $x=y=0.55$ and $\dot{x}=0$ in the Newtonian problem $L_0$. Panel (b) Energy error
$\Delta E= H_0-\tilde{H}_0$, where $H_0$ and $\tilde{H}_0$ are respectively energies at times $t$ and 0.}} \label{fig1}
\end{figure*}

\begin{figure*}
\center{
\includegraphics[scale=0.48]{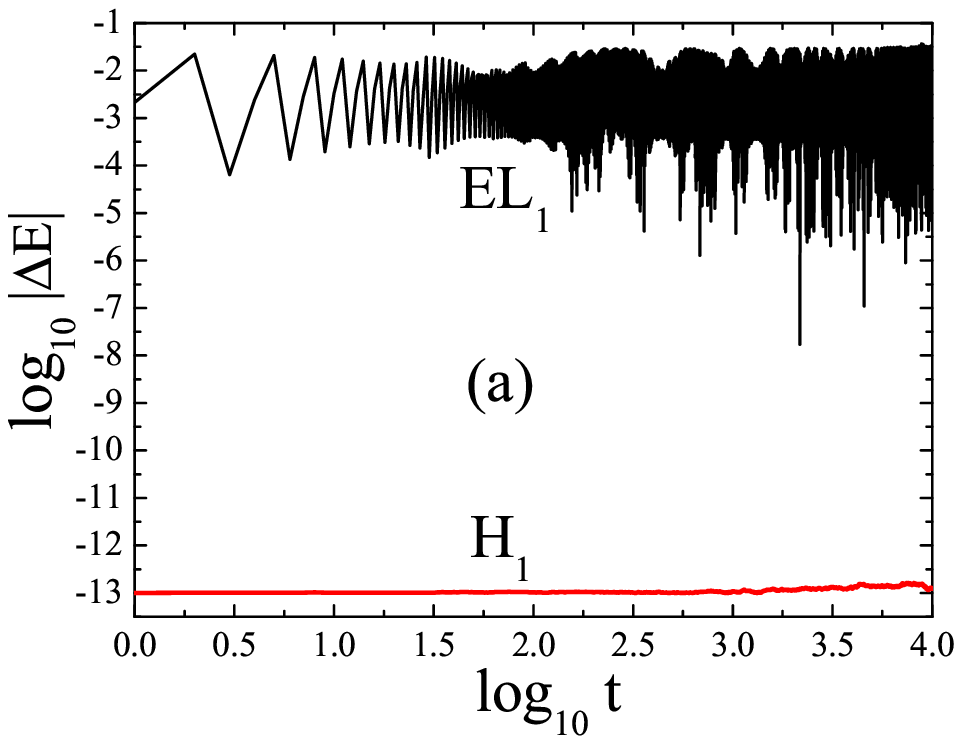}
\includegraphics[scale=0.48]{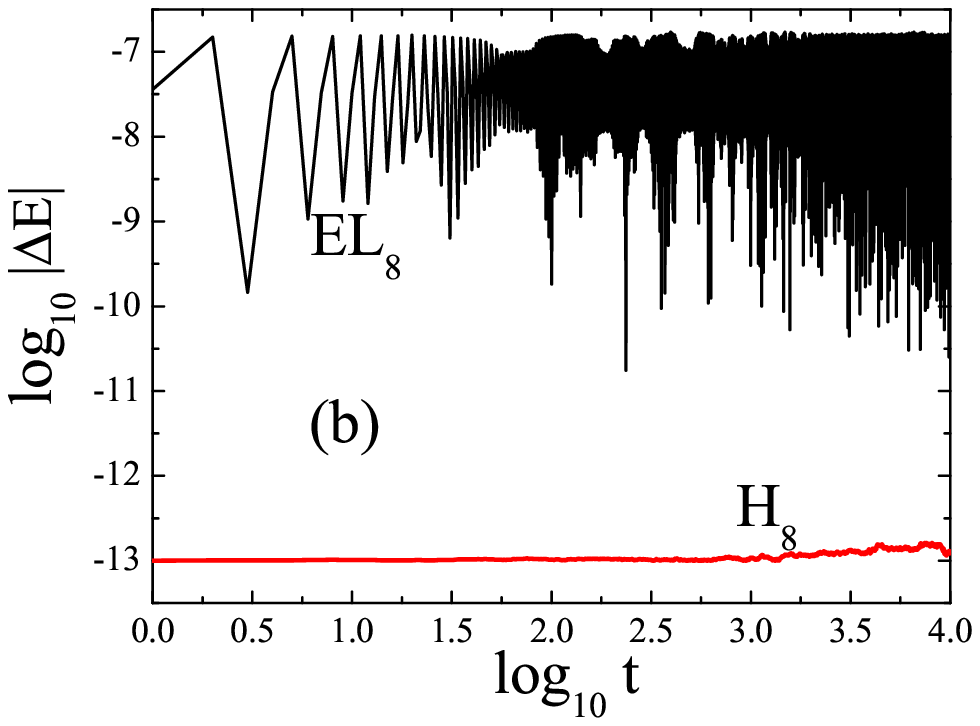}
\includegraphics[scale=0.48]{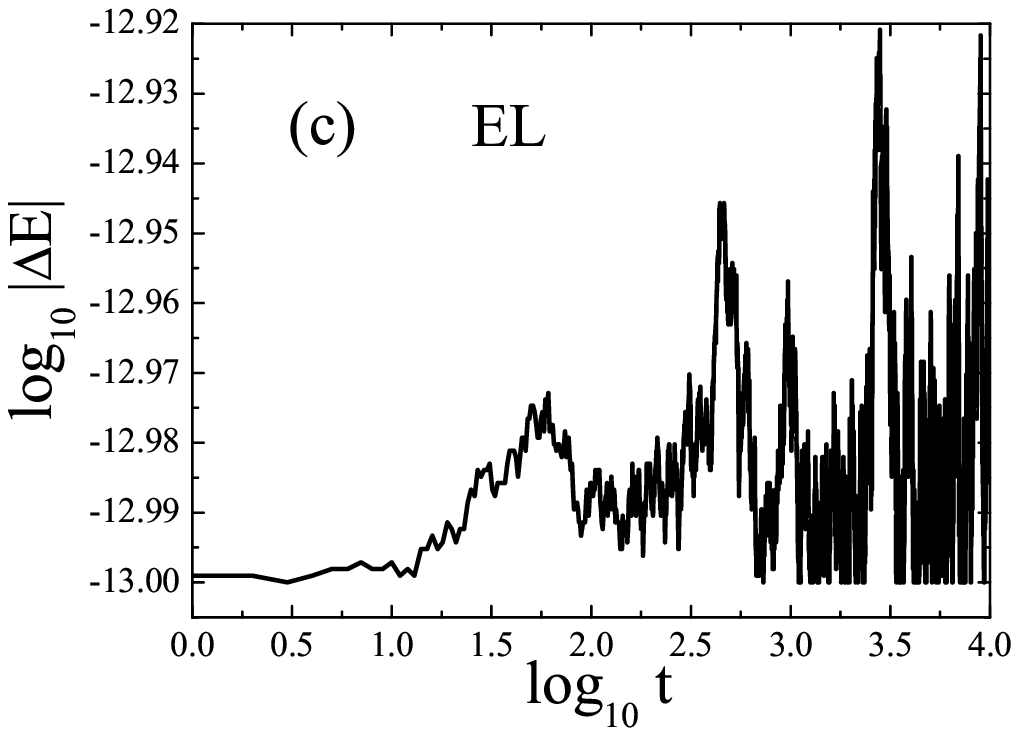}
\caption{(color online) Energy errors
$\Delta E$ for the related PN Lagrangian formulations with the separation $a=31$. Here are some examples to illustrate notations.
For $EL_1$, $\Delta E= H_1-\tilde{H}_1$, where
$\tilde{H}_1$ is the initial energy and the energy $H_1$ at time $t$ is obtained from the solution of $EL_1$.
For $H_1$, $\Delta E= H_1-\tilde{H}_1$, where the energy $H_1$ at time $t$ is obtained from the solution of $H_1$.
For $EL$, $\Delta E= H-\tilde{H}$, where $\tilde{H}$ is the initial energy and the energy $H$ at time $t$ is obtained from the solution of $EL$.
}} \label{fig2}
\end{figure*}

\begin{figure*}
\center{
\includegraphics[scale=0.48]{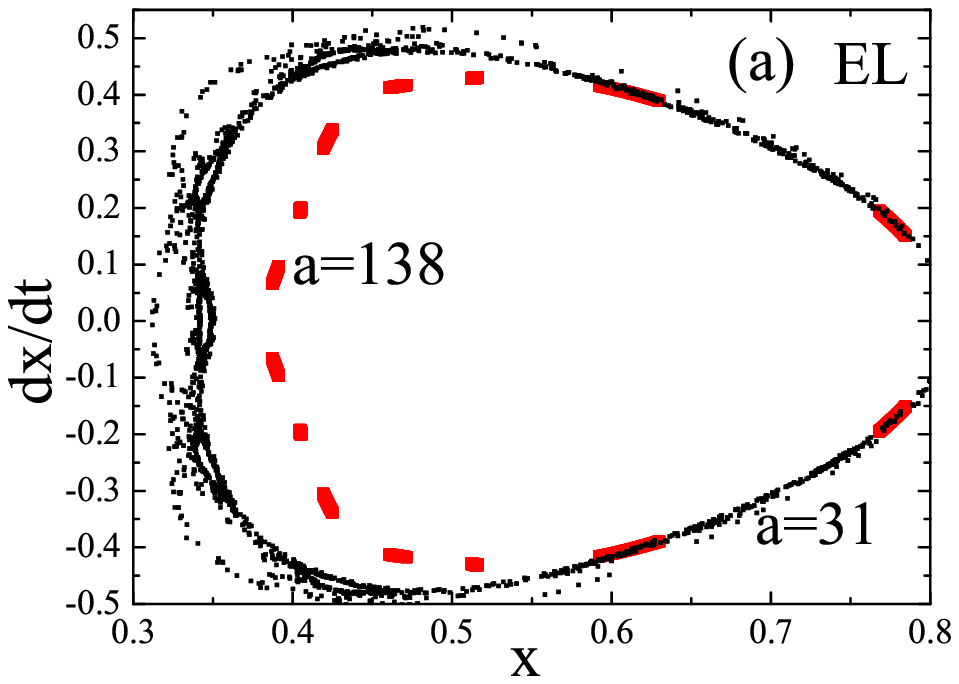}
\includegraphics[scale=0.48]{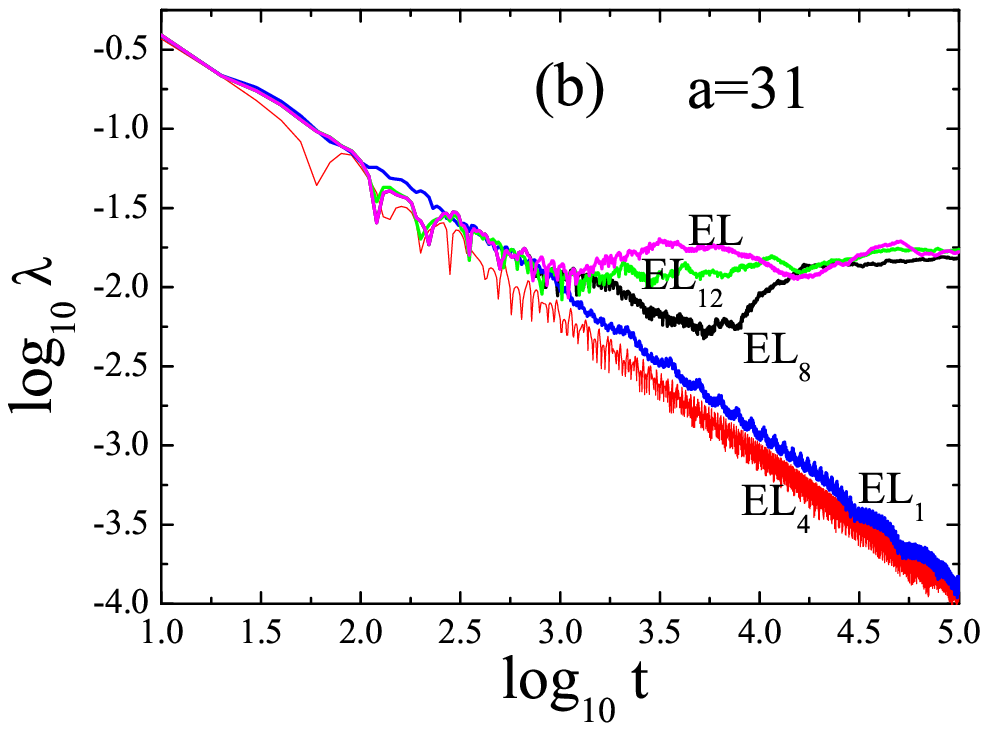}
\includegraphics[scale=0.48]{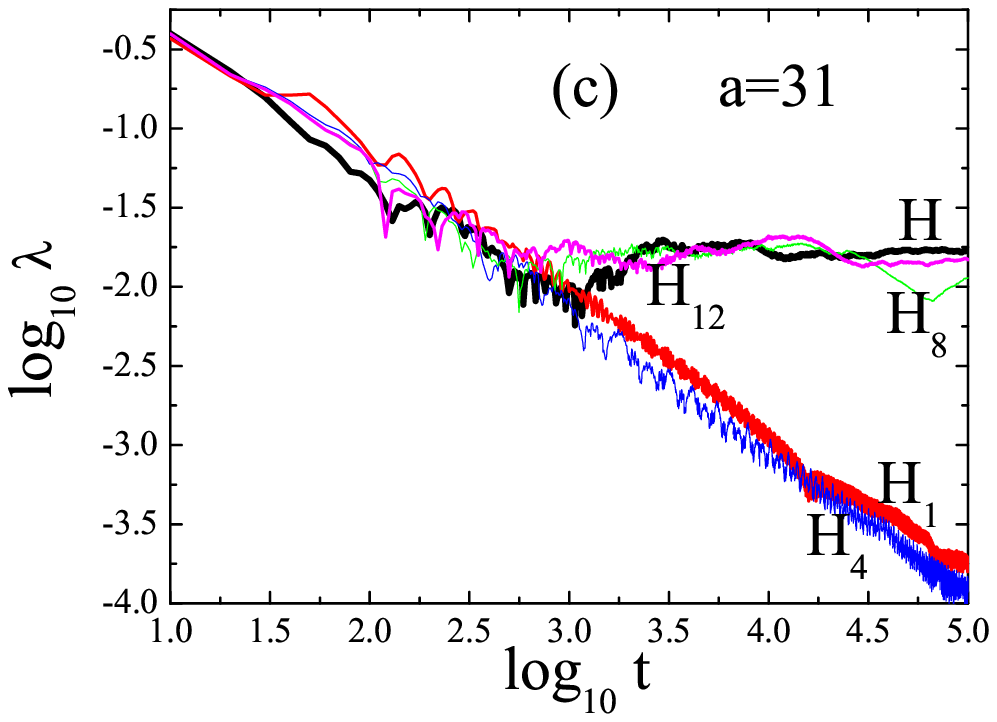}
\includegraphics[scale=0.48]{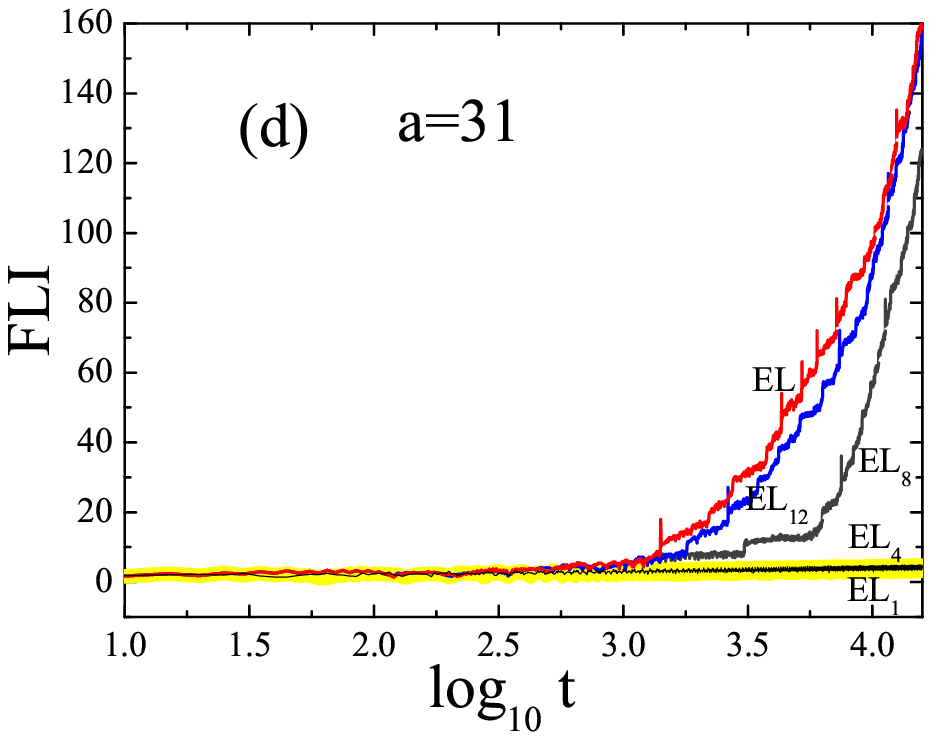}
\includegraphics[scale=0.48]{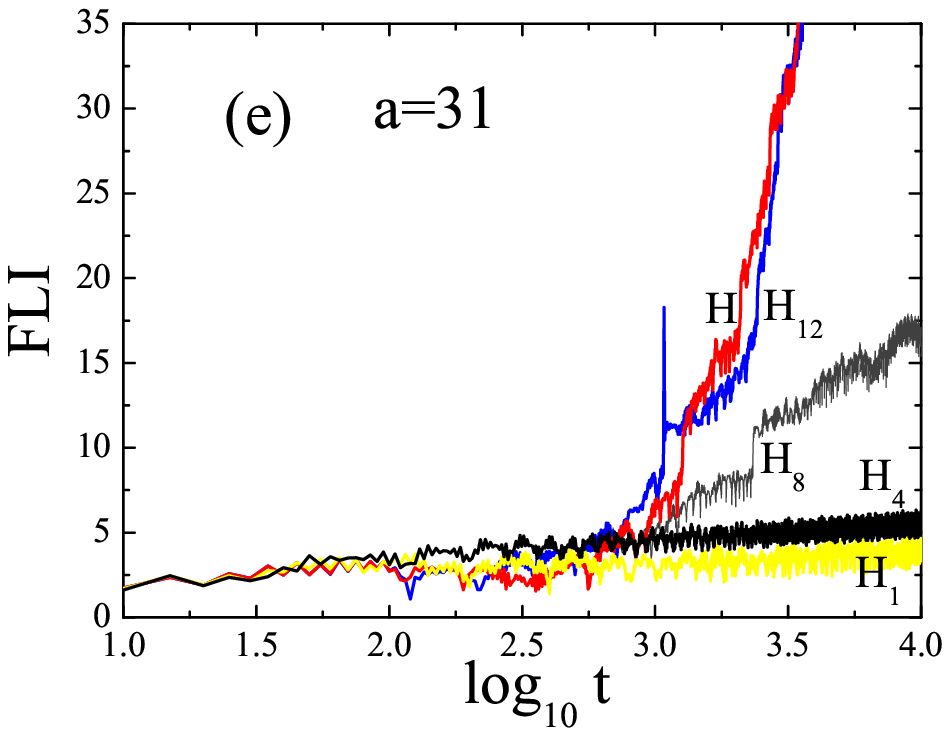}
\includegraphics[scale=0.48]{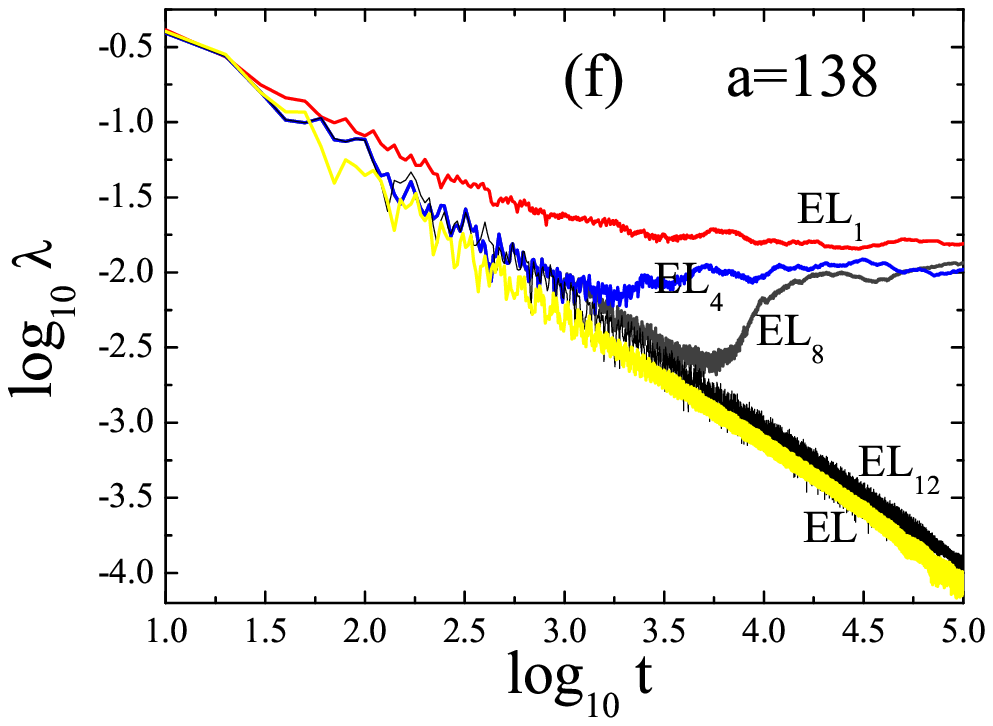}
\includegraphics[scale=0.48]{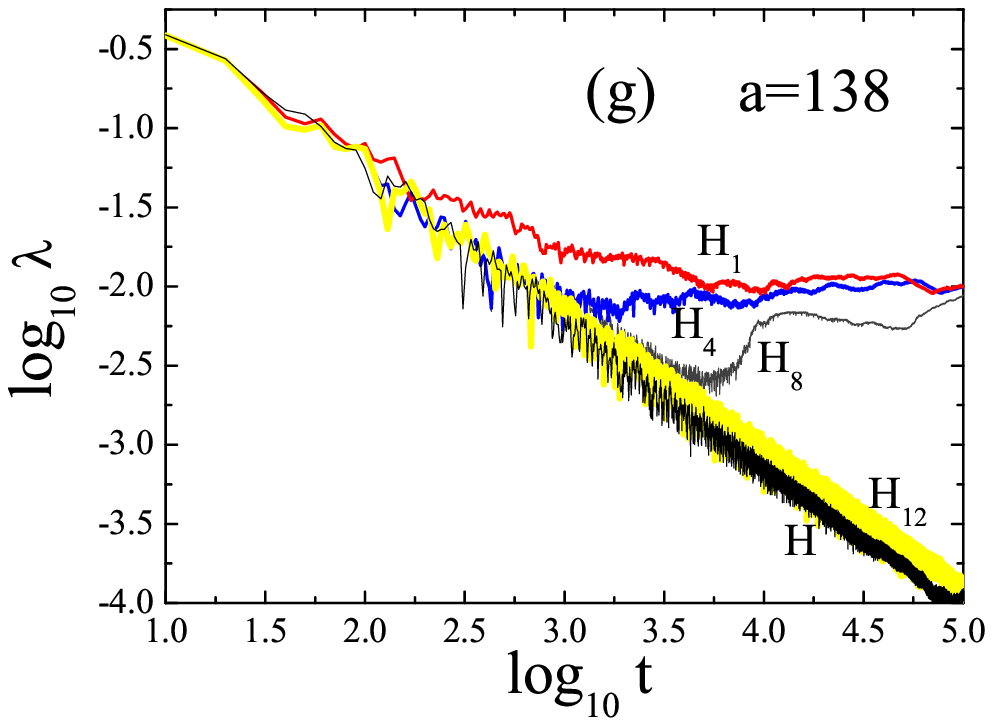}
\includegraphics[scale=0.48]{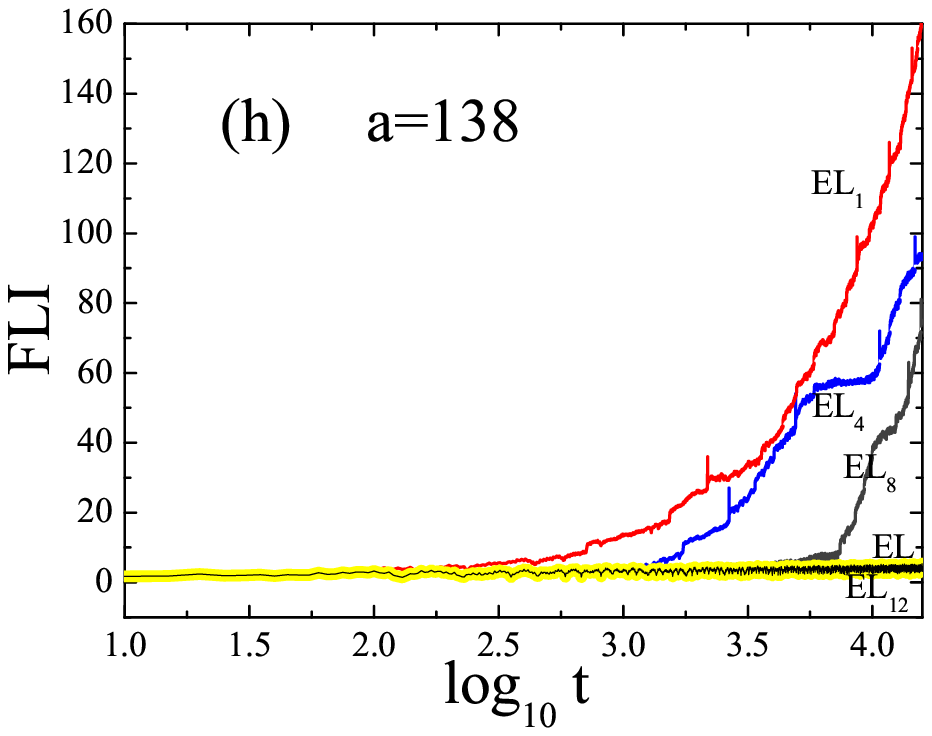}
\includegraphics[scale=0.48]{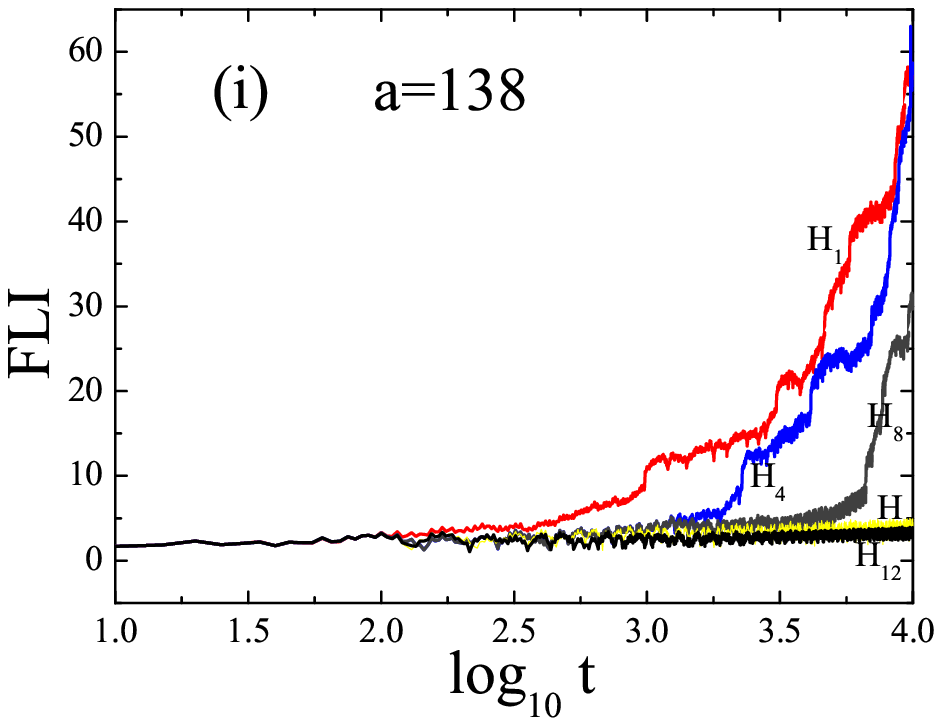}
\caption{(color online) (a) Poincar\'{e} section for the orbit of Fig. 1 in the PN Euler-Lagrange equations $EL$ with
the separation $a=31$ or $a=138$. Panels (b), (c), (f) and (g) relate to Lyapunov exponents $\lambda$, and
panels (d), (e), (h) and (i) deal with FLIs. }} \label{fig3}
\end{figure*}

\begin{figure*}
\center{
\includegraphics[scale=0.68]{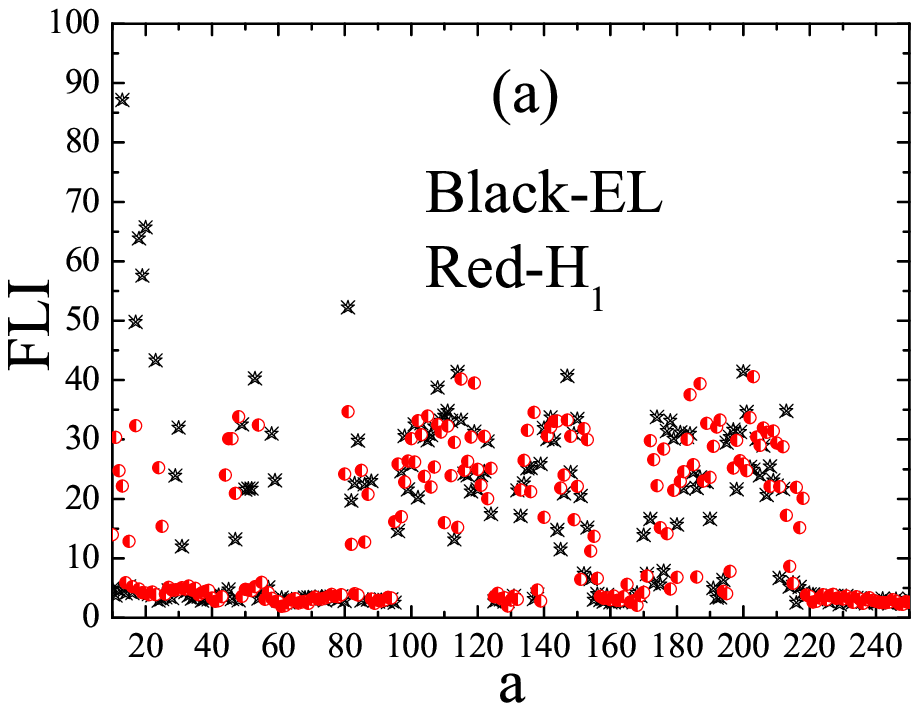}
\includegraphics[scale=0.68]{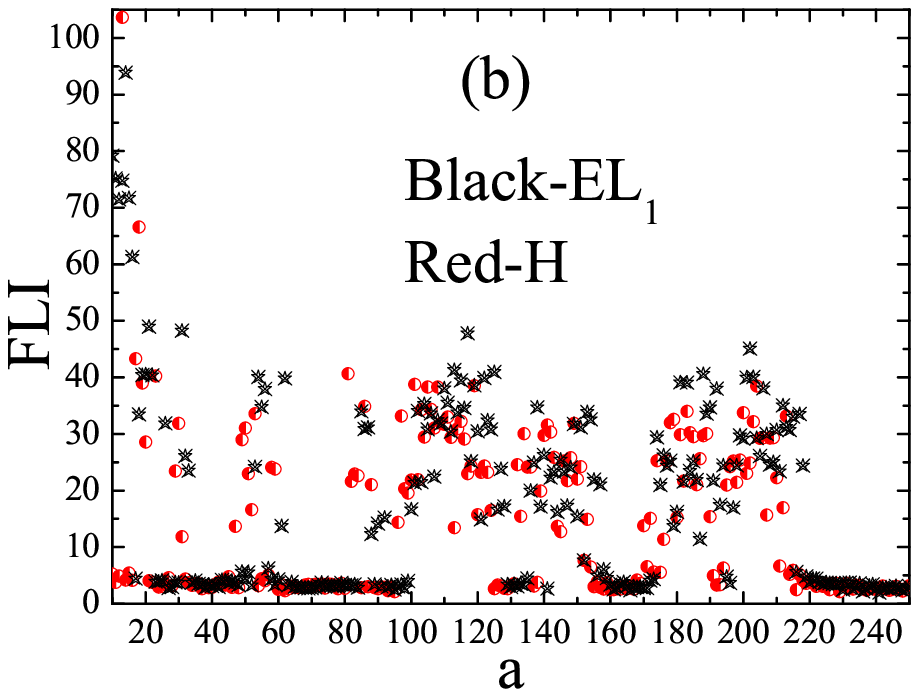}
\includegraphics[scale=0.68]{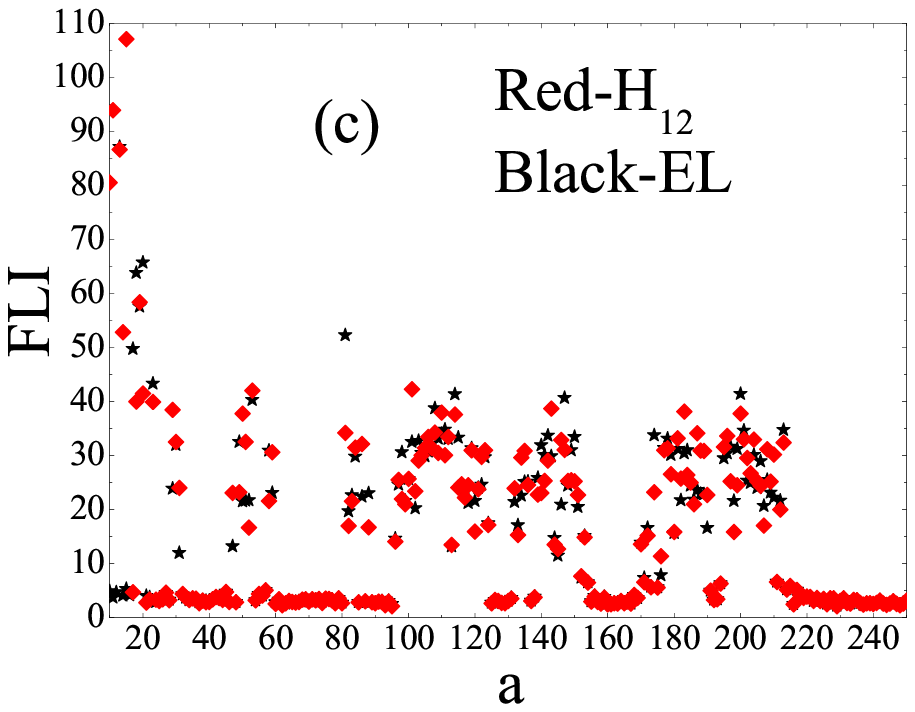}
\includegraphics[scale=0.68]{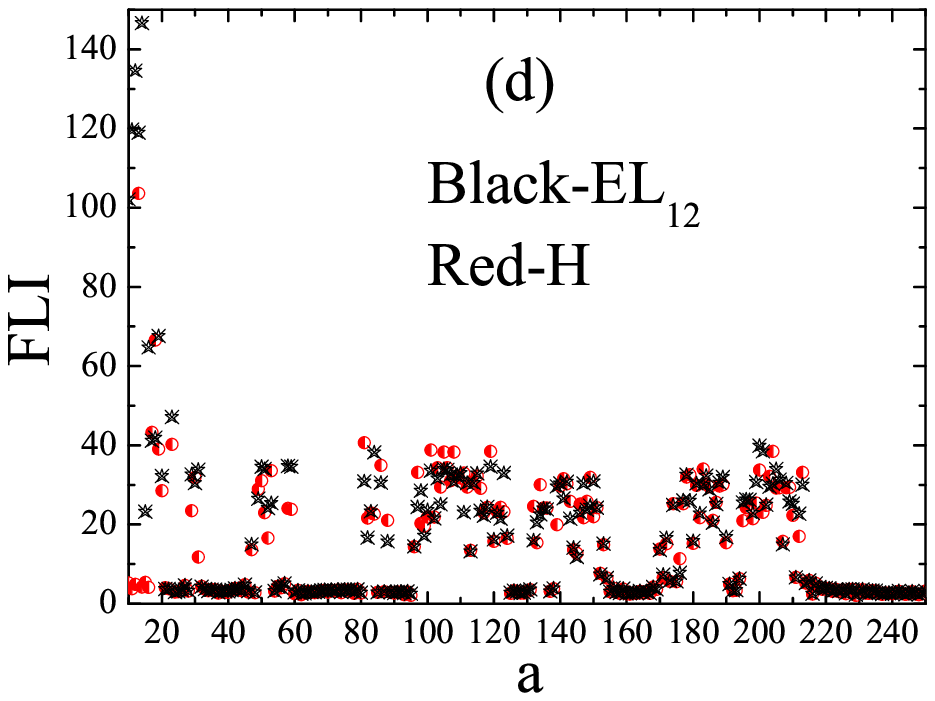}
\caption{(color online) Dependence of FLIs on the separation $a$.}} \label{fig4}
\end{figure*}

\begin{figure*}
\center{
\includegraphics[scale=0.68]{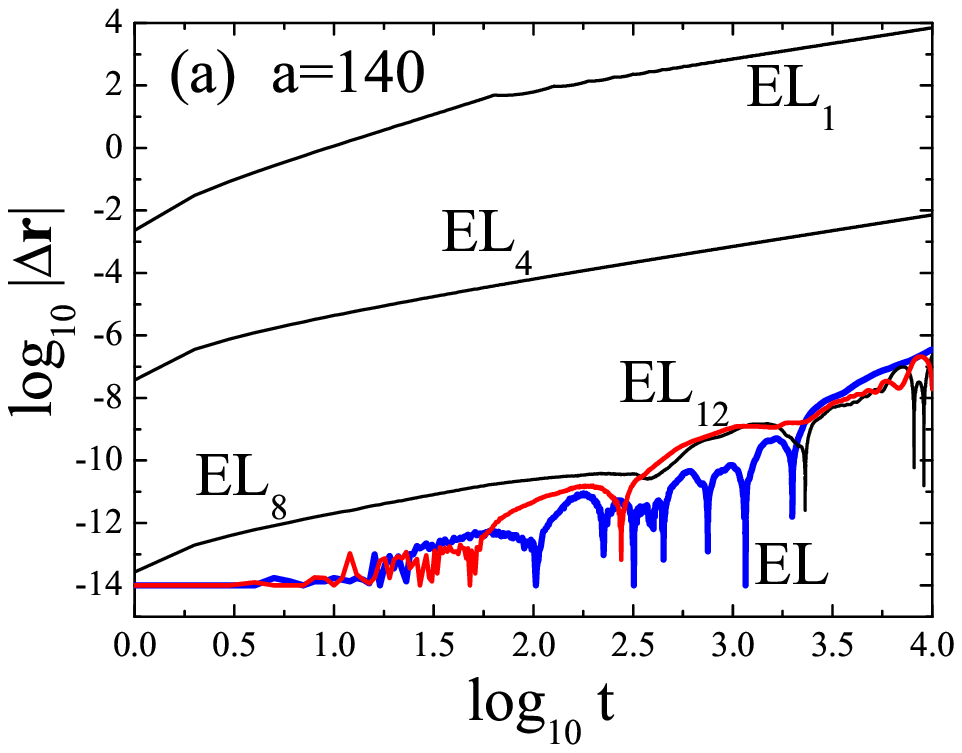}
\includegraphics[scale=0.68]{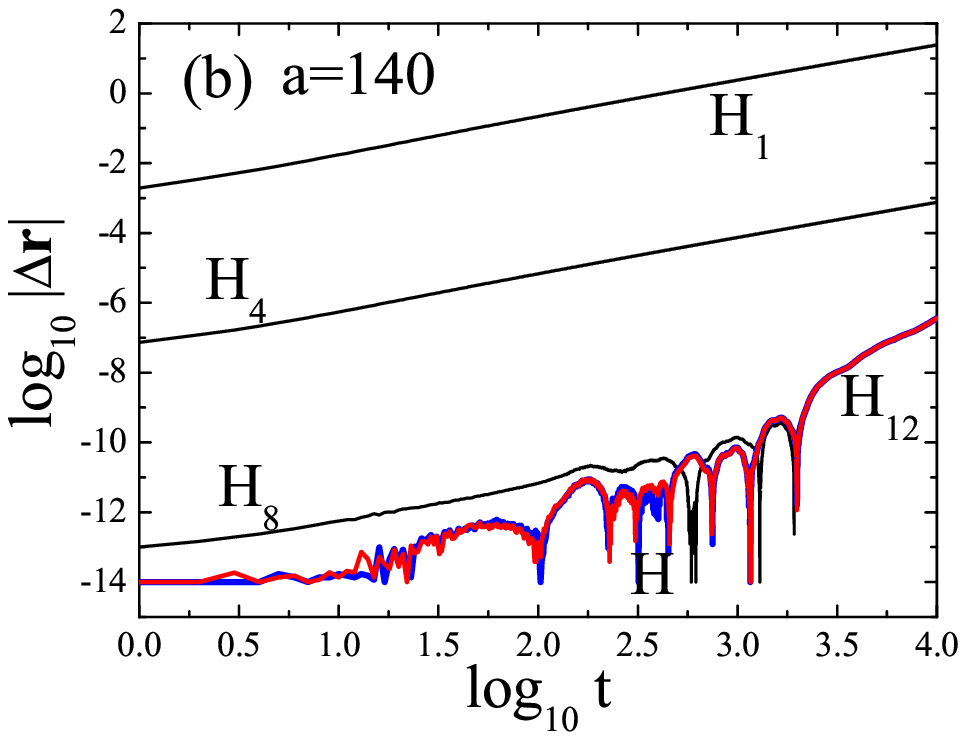}
\includegraphics[scale=0.68]{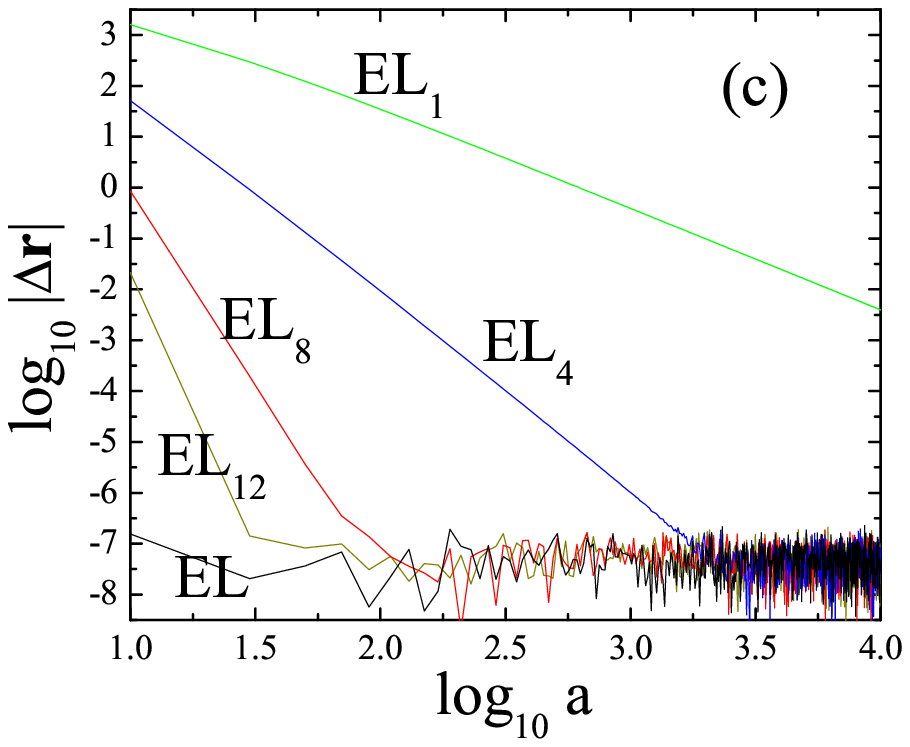}
\includegraphics[scale=0.68]{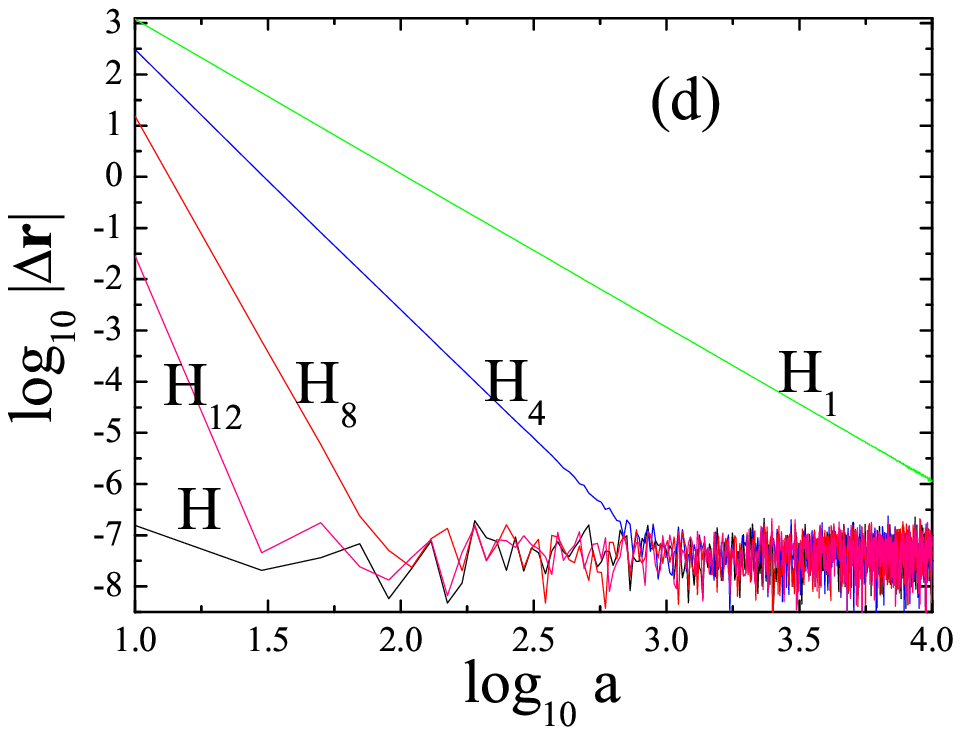}
\caption{(color online) Deviation $|\Delta \mathbf{r}|$ between position solutions of the related
PN Lagrangian and Hamiltonian formulations. Panels (a) and (c) are the deviations from $H$ to $EL$,
$EL_i$ ($i=1,4,8,12$). Panels (b) and (d) deal with the deviations from $EL$ to $H$, $H_i$.}} \label{fig5}
\end{figure*}

\end{document}